\documentclass[letterpaper,twocolumn,showpacs,
prb
,superscriptaddress,email
]{revtex4}

\usepackage[pdftex]{graphicx}  
\usepackage{amssymb}
\usepackage{amsmath,amsfonts,latexsym}
\usepackage{array,tabularx,color}
\usepackage{dcolumn}               
\usepackage[normalem]{ulem}
\usepackage{comment}
\usepackage{bm}        
\usepackage[latin1]{inputenc}
\usepackage{color}

\newcommand{\vect}[1]{{\mathbf #1}}
\newcommand{\Frac}[2]{\displaystyle\frac{#1}{#2}}

\begin{document}

\title{Superflow of resonantly driven polaritons against a defect}

\author{E.~Cancellieri}
\email{emiliano.cancellieri@uam.es}
\affiliation{F\'{\i}sica Te\'orica de la Materia Condensada,
  Universidad Aut\'onoma de Madrid, 28049, Spain.}

\author{F. M. Marchetti}
\affiliation{F\'{\i}sica Te\'orica de la Materia Condensada,
  Universidad Aut\'onoma de Madrid, 28049, Spain.}

\author{M. H. Szyma\'nska}
\affiliation{Department of Physics, University of Warwick, Coventry,
  England.}
\altaffiliation{also at London Centre for Nanotechnology, UK}

\author{C. Tejedor}
\affiliation{F\'{\i}sica Te\'orica de la Materia Condensada,
  Universidad Aut\'onoma de Madrid, 28049, Spain.}

\date{January 11, 2011}       

\begin{abstract}
  In the linear response approximation, coherently driven microcavity
  polaritons in the pump-only configuration are expected to satisfy
  the Landau criterion for superfluidity at either strong enough pump
  powers or small flow velocities. Here, we solve non-perturbatively
  the time dependent Gross-Pitaevskii equation describing the
  resonantly-driven polariton system. We show that, even in the limit
  of asymptotically large densities, where in linear response
  approximation the system satisfies the Landau criterion, the fluid
  always experiences a residual drag force when flowing through the
  defect. We illustrate the result in terms of the polariton lifetime
  being finite, finding that the equilibrium limit of zero drag can
  only be recovered in the case of perfect microcavities. In general,
  both the drag force exerted by the defect on the fluid, as well as
  the height of Cherenkov radiation, and the percentage of particles
  scattered by the defect, show a smooth crossover rather than a sharp
  threshold-like behaviour typical of superfluids which obey the
  Landau criterion.
\end{abstract}






\maketitle

\section{Introduction}
\label{sec:intro}
In the past two decades, microcavity polaritons~\cite{weisbuch92} have
attracted considerable interest because of the possibility of strongly
coupling light and matter, leading to an easy manipulation and
detection of the non-linear properties of matter via light --- see,
e.g., Ref.~\cite{keeling_review07} and references therein. Recently,
in light of their out-of-equilibrium nature, there has been a growing
interest in studying polariton superfluid
properties~\cite{keeling09}. In fact, polaritons have very short
lifetimes (around $\sim 30$~ps even in the best available
samples~\cite{wertz10}), therefore any polariton fluid is the result
of a steady state balance between pumping and decay. The effects of
pump and decay in polariton systems are the subject of several recent
theoretical and experimental work.

For condensates in local thermal equilibrium, such as superfluid
$^4$He and the ultracold atom Bose-Einstein condensates, the concept
of superfluidity is strongly linked to several paradigmatical
properties, such as the Landau criterion, quantised circulation of
velocity, and metastable persistent flow. In particular, the Landau
criterion connects the frictionless motion of a defect at velocities
smaller than a critical one, with the shape of the spectrum of
elementary excitations~\cite{pitaevskii03}. For weakly interacting
Bose gases and a microscopic weak defect, such that a perturbative
linear response theory can be applied, the phonon-like dispersion of
the Bogoliubov spectrum demands that the critical velocity for
dissipationless flow coincides with the velocity of
sound~\cite{astrakharchik04}. However, for macroscopic defects, the
critical velocity for the onset of drag is smaller than the speed of
sound and most likely related to vortex nucleation~\cite{onofrio00}.

In contrast, the relevance of the Landau criterion for
out-of-equilibrium condensates is questionable, not least because the
spectrum of excitations is now complex rather than real. For polariton
fluids, one has to singularly assess the system properties in the
three different pumping schemes available: i) non-resonant pumping;
ii) parametric drive in the optical-parametric-oscillator (OPO)
regime; iii) coherent drive in the pump-only configuration.

As far as the spectrum of quasi-particle excitations and the Landau
criterion are concerned, the cases of non-resonantly pumped
polaritons~\cite{kasprzak06:nature} and parametrically driven
polaritons in the OPO regime~\cite{savvidis00:prl} are similar: in
both cases, there is a $U(1)$ phase symmetry which is spontaneously
broken above a pump power threshold. This leads to the appearance of a
gapless (Goldstone) mode in the spectrum of excitations --- gapless
means that both real and imaginary part of the excitation energy go to
zero at zero momentum. However, in both cases, the effects of pump and
decay are such that the real part of the Goldstone mode energy is zero
also in a finite interval at small momenta, i.e. the spectrum becomes
diffusive~\cite{szymanska06:prl,wouters07,wouters06b}. This means that
a strict application of the Landau criterion would lead to a zero
critical velocity, where quasi-particles can be excited at any value
of the fluid speed.  Therefore, if one would define superfluid
properties through a strict application of the Landau criterion,
neither non-resonantly pumped polaritons nor parametrically driven
polaritons in the OPO regime behave as superfluids. Nevertheless, in
both cases there have been evidences for superfluid behaviour. For
non-resonantly pumped polaritons, it has been recently
shown~\cite{wouters10} that, even though strictly speaking there
cannot be superfluid behaviour, there are regimes close to
equilibrium, where the drag force exerted on a small moving defect
shows a sharp threshold at velocities close to the speed of
sound. Otherwise, for shorter polariton lifetimes, the threshold-like
behaviour of the drag force is replaced by a smooth
crossover. Moreover, metastability of supercurrents in non-resonantly
pumped microcavities has been theoretically
demonstrated~\cite{wouters09}. For polaritons in the OPO regime,
superfluidity has instead been tested through frictionless flow of
polariton bullets~\cite{amo09}, through metastability of quantum
vortices and persistence of currents~\cite{sanvitto09,marchetti10}.

The case of coherently driven polaritons in the pump-only
configuration strongly differs from the two schemes previously
described: Here, there is no phase freedom any longer, because the
polariton phase locks to the one of the driving pump. As a
consequence, the quasi-particle excitation spectrum is always gapped
--- i.e., even when the real part of the spectrum energy goes to zero
at the pump momentum, the corresponding imaginary part is non
zero. For coherently driven polaritons in the pump-only configuration,
by making use of a linearised Bogoliubov-like theory, it has been
predicted that the Landau criterion can instead be satisfied at either
strong enough pump powers or small flow
velocities~\cite{carusotto04,ciuti05}, regimes where polaritons
display a dramatic reduction in the intensity of resonant Rayleigh
scattering. For values of the parameters where instead the spectrum
allows the excitation of quasi-particles, Cherenkov-like waves were
predicted~\cite{ciuti05}, and recently observed in the density
profile~\cite{amo09_b}. Consequently, experiments in this
configuration have been analysed in terms of the same theoretical
description which is valid for equilibrium superfluids.

In this work, we show that, despite the fact coherently driven
polaritons in the pump-only configuration do satisfy the Landau %
criterion at large enough densities or small flow velocities, because
of the polariton lifetime being finite, the fluid always experiences a
residual drag force even in the limit of asymptotically large
densities. We show that, only in the limit of perfect microcavities
(i.e. infinitely long polariton lifetimes), the residual drag force at
large enough densities goes to zero, recovering therefore the
equilibrium limit. Otherwise, for finite polariton lifetimes close to
the current experimental values, we find that, similarly to the case
of incoherently driven polaritons, both the drag force exerted on the
polariton fluid by a defect, as well as the height of Cherenkov
radiation, and the percentage of particles scattered by the defect
show a smooth crossover rather than a sharp threshold-like behaviour
which is typical of superfluids which obey the Landau criterion.

The paper is organised as follows: in Sec.~\ref{sec:model} we
introduce the model describing polaritons coherently driven in the
pump-only configuration in presence of a defect potential. In
Sec.~\ref{sec:metho} we describe the methods we use for our analysis,
in particular the numerical algorithm used to evaluate quantities such
as the drag force, the height of Cherenkov radiation, and the
percentage of particles scattered by the defect which characterise the
crossover from a superfluid-like to a supersonic-like behaviour. In
addition, in Sec.~\ref{sec:linea}, we shortly introduce the linearised
Bogoliubov-like theory of Refs.~\cite{carusotto04,ciuti05}, which will
be used later in Sec.~\ref{sec:resul} in order to compare the results
obtained with the non-perturbative method with the results obtained in
the linear response approximation. Results are discussed in
Sec.~\ref{sec:resul}, while conclusions (together with a discussion of
the experimental relevance of our findings) are drawn in
Sec.~\ref{sec:concl}.

\section{Model}
\label{sec:model}
%
We describe the dynamics of the resonantly-driven polariton
system~\cite{ciuti03} via a Gross-Pitaevskii equation for coupled
cavity and exciton fields $\psi_{C,X} (\vect{r},t)$, generalised to
include the effects of the resonant pumping and decay ($\hbar=1$):
\begin{widetext}
\begin{equation}
  i\partial_t \begin{pmatrix} \psi_X \\ \psi_C \end{pmatrix}
  = \begin{pmatrix} 0 \\ F_p \end{pmatrix} +
  \left[\hat{H}_0 (-i\nabla) + \begin{pmatrix} -i \kappa_X + g_X|\psi_X|^2& 0 \\ 0 & -i
  \kappa_C + V_d (\vect{r}) \end{pmatrix}\right] \begin{pmatrix}
    \psi_X \\ \psi_C
  \end{pmatrix}\; .
\label{eq:model}
\end{equation}
\end{widetext}
The single-particle polariton Hamiltonian $\hat{H}_0$ can be
diagonalised in momentum space,
\begin{equation}
  \hat{H}_0(\vect{k}) = \begin{pmatrix} \omega_{X} (\vect{k}) &
    \Omega_R/2 \\ \Omega_R/2 & \omega_{C}(\vect{k})\end{pmatrix} \; ,
\end{equation}
by rotating into the lower (LP) and upper polariton (UP) basis,
\begin{equation*}
  \begin{pmatrix} \psi_X \\ \psi_C \end{pmatrix} = \begin{pmatrix}
    \cos \theta_{\vect{k}} & -\sin \theta_{\vect{k}} \\ \sin
    \theta_{\vect{k}} & \cos
    \theta_{\vect{k}}\end{pmatrix} \begin{pmatrix} \psi_{LP}
    \\ \psi_{UP}
    \end{pmatrix} \; ,
\end{equation*}
where $\tan 2\theta_{\vect{k}} =
\Omega_R/[\omega_{X}(\vect{k})-\omega_{C}(\vect{k})]$, giving the bare
lower and upper polariton dispersions:
\begin{equation}
  \omega_{LP,UP}(\vect{k}) = \Frac{\omega_{X} + \omega_{C}}{2} \mp
  \Frac{\sqrt{[\omega_{C} - \omega_{X}]^2 + \Omega_R^2}}{2}\; .
\label{eq:bared}
\end{equation}
At zero energy detuning between excitons and photons, $\omega_{C}
(\vect{k}) = \omega_{X} (\vect{k})$, lower and upper polaritons are an
equally balanced mixture of exciton and photon, i.e. $\sin^2
\theta_{\vect{k}} = \cos^2 \theta_{\vect{k}} = 1/2$. In contrast, at
large values of the energy detuning between photons and excitons,
$\omega_C (\vect{k}) - \omega_X (\vect{k}) \gg \Omega_R$,
i.e. $\theta_{\vect{k}} \to 0$, the lower and upper polariton
respectively coincide with the exciton and the photon.

In the following, we will neglect the exciton dispersion,
$\omega_X(\vect{k}) = \omega_X (0)$, and assume a quadratic dispersion
for photons, $\omega_C (\vect{k})=\omega_C (0) + \frac{k^2}{2m_C}$,
where the photon mass is $m_C=2\times10^{-5}m_0$ and $m_0$ is the bare
electron mass. Through the paper, we will consider the case of zero
detuning at normal incidence, $\omega_X (0)=\omega_C(0)$, and fix the
Rabi frequency to $\Omega_R=4.4$~meV. The parameters $\kappa_X$ and
$\kappa_C$ are respectively the excitonic and photonic decay rates. We
will fix these parameters in order to give a polariton lifetime,
$\tau_{LP} = \hbar/\kappa_{LP}$,
\begin{equation}
  \kappa_{LP} (\vect{k}) = \kappa_X \cos^2 \theta_{\vect{k}} +
  \kappa_C \sin^2 \theta_{\vect{k}} \; ,
\label{eq:lifet}
\end{equation}
close to the experimental values. In addition, we will consider the
limit of perfect cavities $\kappa_{LP} \to 0$ in order to recover the
equilibrium limit.

In Eq.~\eqref{eq:model}, the exciton-exciton interaction strength
$g_X$ can be set to one by rescaling both fields $\psi_{X,C}$ and pump
strength $F_{p}$ by $\sqrt{\Omega_R/(2 g_X)}$. The cavity field is
driven by a continuous-wave pump,
\begin{equation}
  F_p(\vect{r},t) = \mathcal{F}_{f,\sigma} (r) e^{i (\vect{k}_p \cdot
    \vect{r} - \omega_p t)}\; ,
\label{eq:fpump}
\end{equation}
with a smoothen top-hat spatial profile with intensity $f$ and full
width at half maximum (FWHM) $\sigma=130~\mu$m. The pumping laser
frequency $\omega_p$ has been chosen $0.44$~meV blue-detuned above the
bare lower polariton dispersion at the pump momentum $\omega_{LP}
(\vect{k}_p)$.  The polariton flow current is determined by the pump
momentum $\vect{k}_p$, which can be experimentally tuned by changing
the pumping laser angle of incidence with respect to the growth
direction $\varphi_{\vect{k}_p}$:
\begin{equation*}
  c \vect{k}_p = \omega_{LP} (\vect{k}_p) \sin(\varphi_{\vect{k}_p})\;
  .
\end{equation*}
Finally, in Eq.~\eqref{eq:model} the potential $V_d (\vect{r})$
describes the defect acting on the photonic field, over which the
polariton fluid scatters. Specifically, we consider:
\begin{equation}
  V_d (\vect{r})= V_d \theta(r - r_d)\; ,
\label{eq:defec}
\end{equation}
with $r_d=7~\mu$m and $V_d=110$~meV. Defects can be present naturally
in the sample's mirror~\cite{amo09_b}. Alternatively, defects can been
artificially engineered by either growing mesas in one of the
mirrors~\cite{eldaif06} or by an additional laser~\cite{amo10}.

\section{Methods}
\label{sec:metho}
%
We numerically solve Eq.~\eqref{eq:model} on a 2D grid
($256\times256$) in a $150~\mu$m$\times$$150~\mu$m box using a
5$^{\text{th}}$-order adaptive-step Runge-Kutta algorithm, and
evaluate both exciton and photon wave-functions $\psi_{X,C}
(\vect{r},t)$ in the steady-state regime.

\begin{figure}
\begin{center}
\includegraphics[width=1.0\linewidth,angle=0]{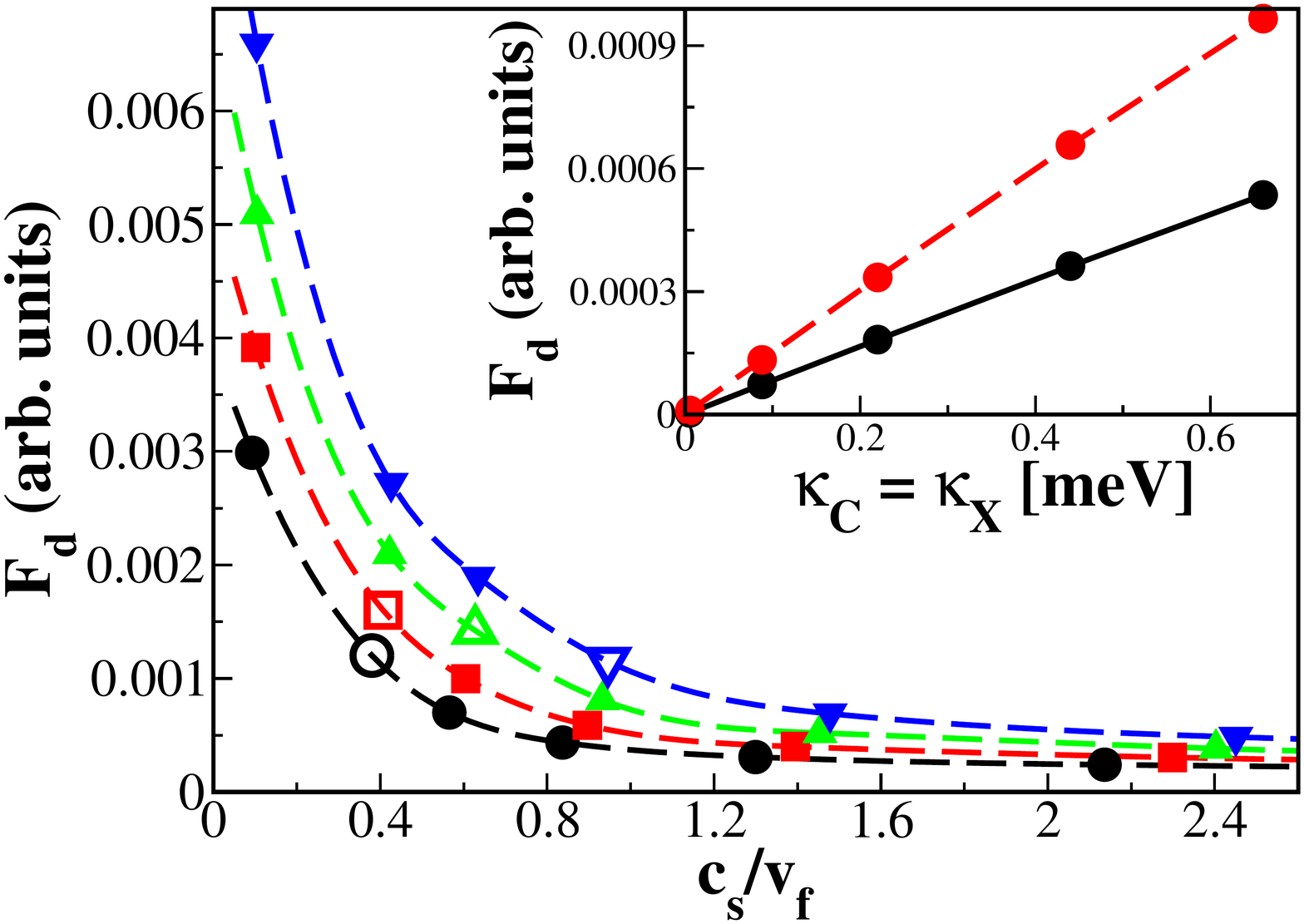}
\end{center}
\caption{(Color online) Drag force in the $x$-direction,
  ${F_d}_x$~\eqref{eq:dragf}, exerted on the polariton fluid by the
  defect~\eqref{eq:defec} as a function of the velocity ratio
  $c_s/v_f$ (see Eqs.~\eqref{eq:fluid} and~\eqref{eq:sound}), where
  $n_{LP} = \int d\vect{r}|\psi_{LP} (\vect{r})|^2/\Omega$ is the
  average polariton density in the area $\Omega$, and for different
  values of the pump momentum: $k_p=0.7~\mu$m$^{-1}$ (black circles),
  $k_p=0.8~\mu$m$^{-1}$ (red squares), $k_p=0.9~\mu$m$^{-1}$ (green up
  triangles), and $k_p=1~\mu$m$^{-1}$ (blue down triangles). The
  exciton and photon decay rates have been fixed to $\kappa_X=
  \kappa_C = 0.22$~meV, while the pumping laser frequency $\omega_p$
  has been chosen (as in all the Figures) $0.44$~meV blue-detuned
  above the bare LP dispersion $\omega_{LP} (\vect{k}_p)$. The empty
  symbols indicate the value of $c_s/v_f$ above which in the linear
  approximation the Landau condition of Eq.~\eqref{eq:landa} cannot be
  satisfied and therefore where in principle one expects the drag
  force going to zero. Inset: Residual drag force at asymptotically
  large polariton densities $n_{LP}$ (i.e., $c_s/v_f \to \infty$) as a
  function of the exciton and photon decay rates for
  $k_p=0.7~\mu$m$^{-1}$ (black solid line) and $k_p=1~\mu$m$^{-1}$
  (red dashed line).}
\label{fig:dragf}
\end{figure}

We characterise the crossover from a superfluid-like to a
supersonic-like regime by evaluating three different
quantities. Firstly, we consider the normalised drag
force~\cite{astrakharchik04} exherted on the flowing polaritons by the
defect~\eqref{eq:defec}:
\begin{equation}
  \vect{F}_d = \frac{1}{\int d\vect{r} |\psi_C(\vect{r})|^2} \int
  d\vect{r} |\psi_C(\vect{r})|^2 \nabla V_d (\vect{r})\; .
\label{eq:dragf}
\end{equation}
When shining the laser pump in the $x$-direction,
$\vect{k}_p=(k_p,0)$, the density profile will be symmetric under the
transformation $y \mapsto -y$, $\psi_{X,C}(x,y) = \psi_{X,C}(x,-y)$,
implying that for defect potentials symmetric under $y \mapsto -y$
only the $x$-component of the drag force can be non-zero. Moreover,
for step-like defects such as~\eqref{eq:defec}, only the values of the
field $|\psi_C(\vect{r})|^2$ at a distance $r=r_d$ contribute to the
integral~\eqref{eq:dragf}:
\begin{equation}
  {F_d}_x = \frac{2V_d r_d }{\int d\vect{r} |\psi_C(\vect{r})|^2}
  \int_0^{\pi} d\phi \cos\phi |\psi_C(r_d,\phi)|^2 \; .
\label{eq:asymm}
\end{equation}
Therefore the drag force measures the degree of asymmetry of the
photon (or alternatively the exciton) density profiles going from a
certain angle $\phi<\pi/2$ ahead (with respect to the fluid flow
direction) of the defect, to an angle $\pi - \phi>\pi/2$ behind the
defect. The asymmetry is caused by the scattering of the fluid passing
through the defect. Plots of the drag force are drawn in
Fig.~\ref{fig:dragf}.

Second, we characterise the superfluid-like behaviour by the
suppression of density modulations around the defect known as
\emph{Cherenkov waves} --- see, e.g.,
Refs.~\cite{carusotto06_prl,gladush07} and references therein for
atomic Bose-Einstein condensates and Refs.~\cite{carusotto04,ciuti05}
for coherently driven polaritons in the pump-only
configuration. Cherenkov radiation is generated in the supersonic
regime, when the fluid is passing a defect at a velocity higher than
the phase velocity of the fluid elementary excitations. As mentioned
later, a simple analysis of Cherenkov radiation can be carried on by
making use of perturbative Bogoliubov-like analysis: Here, in
agreement with the Landau criterion, in the supersonic regime, the
kinetic energy of the fluid can be dissipated radiating Bogoliubov
modes, giving rise to perturbations which propagate radially from the
defect, with a characteristic unperturbed region inside a Cherenkov
cone which is related to the singularity of the Bogoliubov mode
dispersion evaluated in the reference frame of the moving
fluid~\cite{carusotto06_prl}. In our numerical non-perturbative
analysis, we determine the value of the highest crest of the waves,
$h_{\text{max}}$, above the mean value of the fluid,
$h_{\text{mean}}$, giving $h_{\text{{\u C}er}} =
h_{\text{max}}/h_{\text{mean}}$, and analyse how $h_{\text{{\u C}er}}$
changes by tuning the fluid velocities and densities.

Finally, a third way of describing the crossover from a
superfluid-like to a supersonic-like regime is by evaluating the
percentage of particles scattered by the defect with respect to the
total number of particles in the system~\cite{iasenelli06}, quantity
which we label as $S_{out}$. Both $h_{\text{{\u C}er}}$ and $S_{out}$
are plotted in Fig.~\ref{fig:heigh}.

\subsection{Linear response to a weak external potential}
\label{sec:linea}
%
In order to connect our numerical results obtained by solving
non-perturbatively the dynamics of the Gross-Pitaevskii
equation~\eqref{eq:model} with the Landau criterion, and the spectrum
of quasi-particle excitations, similarly to what is done for
conservative weakly interacting Bose gases~\cite{pitaevskii03}, we
follow the same perturbative Bogoliubov-like analysis first introduced
for resonantly pumped polaritons in Refs.~\cite{carusotto04,ciuti05}
and expand both exciton and photon fields above their mean-field
spatially homogeneous and stationary states $e^{ - i (\omega_p t -
  \vect{k}_p \cdot \vect{r})} \psi_{X,C}^{(0)}$:
\begin{equation}
  \psi_{X,C} (\vect{r},t) = e^{ - i \omega_p t} \left[e^{i \vect{k}_p
      \cdot \vect{r}} \psi_{X,C}^{(0)} + \delta \psi_{X,C}
    (\vect{r},t)\right]\; ,
\end{equation}
where we are assuming the pump to have a homogeneous profile
$\mathcal{F}_{f,\sigma} (r) = f$. While the mean-field equations
\begin{align*}
  \left[\omega_X(0) - \omega_p - i\kappa_X + g_X
    |\psi_{X}^{(0)}|^2\right] \psi_{X}^{(0)} + \Frac{\Omega_R}{2}
  \psi_{C}^{(0)} &= 0 \\
  \left[\omega_C(\vect{k}_p) - \omega_p - i\kappa_C\right]
  \psi_{C}^{(0)} + \Frac{\Omega_R}{2} \psi_{X}^{(0)} &= -f
\end{align*}
allow to determine the intensity of exciton and photon fields in
absence of the external potential, fluctuations above mean-field need
to be introduced in order to evaluate the stability of such a
solutions as well as the linear response to a weak perturbing external
potential.

\begin{figure*}
\begin{center}
\includegraphics[width=0.8\linewidth,angle=0]{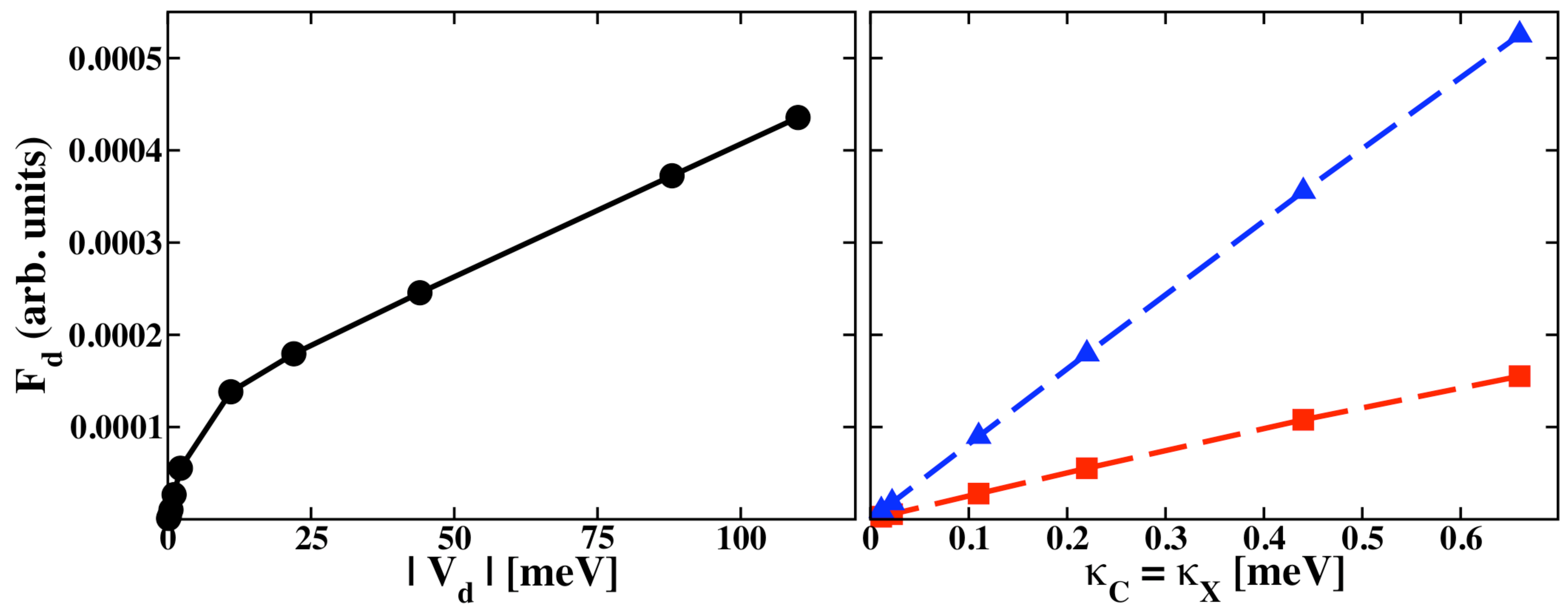}
\end{center}
\caption{(Color online) Residual drag force (in arbitrary units) at
  asymptotically large polariton densities $n_{LP}$ for
  $k_p=0.7~\mu$m$^{-1}$. Left panel gives the dependence on the
  barrier height (in meV) for $\kappa_X=\kappa_C=0.22$~meV.  Right
  panel shows the dependence on the decay rates $\kappa_X$ and
  $\kappa_C$ (in meV) for two different barriers: $V_d=-22$~meV (blue
  dashed line) and $V_d=-2.2$~meV (red dashed line).}
\label{newfig}
\end{figure*}
The spectrum of the quasi-particle excitations can be found by
introducing particle-like $u_{X,C}$ and hole-like $v_{X,C}$
excitations in both the exciton and photon fluctuation fields:
\begin{multline*}
  \delta \psi_{X,C} (\vect{r},t) \\
  = \sum_{\vect{k}} \left(e^{-i \omega t} e^{i \vect{k} \cdot
    \vect{r}} u_{X,C;\vect{k}} + e^{i \omega t} e^{-i (\vect{k} -
    2\vect{k}_p) \cdot \vect{r}} v_{X,C;\vect{k}} \right) \; ,
\end{multline*}
and by solving the eigenvalue equation:
\begin{equation}
  \left[(\omega + \omega_p) \mathbb{I} -
    \mathbb{L}\right] \begin{pmatrix} u_{X;\vect{k}} & u_{C;\vect{k}}
    & v_{X;\vect{k}} & v_{C;\vect{k}}\end{pmatrix}^T = 0 \; ,
\label{eq:spect}
\end{equation}
where $\mathbb{L}$ is a $4\times 4$ matrix given by:
\begin{widetext}
\begin{equation*}
  \mathbb{L} =
\begin{pmatrix}
  \omega_X(0) + 2g_X |\psi_X^{(0)}|^2 -i\kappa_X &
  \Omega_R/2 & g_X {\psi_X^{(0)}}^2 & 0\\
  \Omega_2/2 & \omega_C(\vect{k}) -i\kappa_C & 0 & 0\\
  -g_X {{\psi_X^{(0)}}^2}^* & 0 & 2\omega_p - \omega_X(0) - 2g_X
  |\psi_X^{(0)}|^2 -i\kappa_X & -\Omega_R/2 \\
  0 & 0 & -\Omega_R/2 & 2\omega_p - \omega_C(2\vect{k}_p - \vect{k})
  -i\kappa_C
\end{pmatrix} \; .
\end{equation*}
\end{widetext}
Eq.~\eqref{eq:spect} admits four complex eigenvalues for each
$\vect{k}$ which we indicate as $\omega+\omega_p =
\text{LP}^{\pm}(\vect{k}) , \text{UP}^{\pm} (\vect{k})$. Note that the
spectrum obtained this way is already evaluated in the polariton flow
moving frame. This becomes evident in the limit where the pump
momentum $\vect{k}_p$ is small enough so that the LP dispersion can be
approximated as parabolic, and the UP dispersion can be neglected. In
fact, here, one can show~\cite{ciuti05} that the spectrum of
excitation reduces to
\begin{multline}
  \text{LP}^{\pm}(\vect{k}) - \omega_p \simeq \vect{v}_f \cdot
  (\vect{k} - \vect{k}_p) -i\kappa_{LP} \\
  \pm \sqrt{\left(\varepsilon_{\vect{k}} - \Delta\right)
    \left(\varepsilon_{\vect{k}} - \Delta + 2 g_{LP}
    |\psi_{LP}^{(0)}|^2\right)} \; ,
\label{eq:LPlim}
\end{multline}
where
\begin{equation}
  \vect{v}_f = \Frac{\vect{k}_p}{m_{LP}}
\label{eq:fluid}
\end{equation}
is the polariton fluid velocity at the pump momentum,
$\varepsilon_{\vect{k}} = (\vect{k} - \vect{k}_p)^2/(2 m_{LP})$,
$m_{LP} = m_C/\sin^2 \theta_{\vect{k}_p}$ is the LP mass, $\Delta =
\omega_p - \omega_{LP} (\vect{k}_p) - g_{LP} |\psi_{LP}^{(0)}|^2$ is
the pump detuning renormalised by the interaction, and $g_{LP} = g_X
\cos^4\theta_{\vect{k}_p}$.  For more details, we refer the reader to
the original calculation of Ref.~\cite{ciuti05}. What we would like to
stress here is how to generalise the Landau criterion to the complex
spectrum of elementary excitations obtained from
Eq.~\eqref{eq:spect}. In particular, in the linear approximation, we
expect that the perturbation introduced by the defect is able to
excite Bogoliubov-like quasi-particle states with momentum $\vect{k}$,
when the condition
\begin{equation}
  \Re[\text{LP}^{+} (\vect{k})] - \omega_p< 0
\label{eq:landa}
\end{equation}
is satisfied. Note that, in the limit where the
approximation~\eqref{eq:LPlim} is valid and at resonance, $\Delta =
0$, this recovers the Landau criterion in its original formulation for
a conservative system, i.e. close to $\vect{k}_p$, $\Re[\text{LP}^{+}
  (\vect{k})] - \omega_p \simeq (c_s \pm v_f) |\vect{k} - \vect{k}_p|$
and the critical fluid velocity coincides with the sound velocity
given by the usual~\cite{pitaevskii03} expression:
\begin{equation}
  c_s \equiv \sqrt{\Frac{g_{LP} n_{LP}}{m_{LP}}}\; ,
\label{eq:sound}
\end{equation}
where $n_{LP}=|\psi_{LP}^{(0)}|^2$ is the mean-field polariton
density. In order to be able to draw an analogy with the equilibrium
limit, we will later present our results in terms of the ratio
$c_s/v_f$.

\begin{figure}
\begin{center}
\includegraphics[width=1.0\linewidth,angle=0]{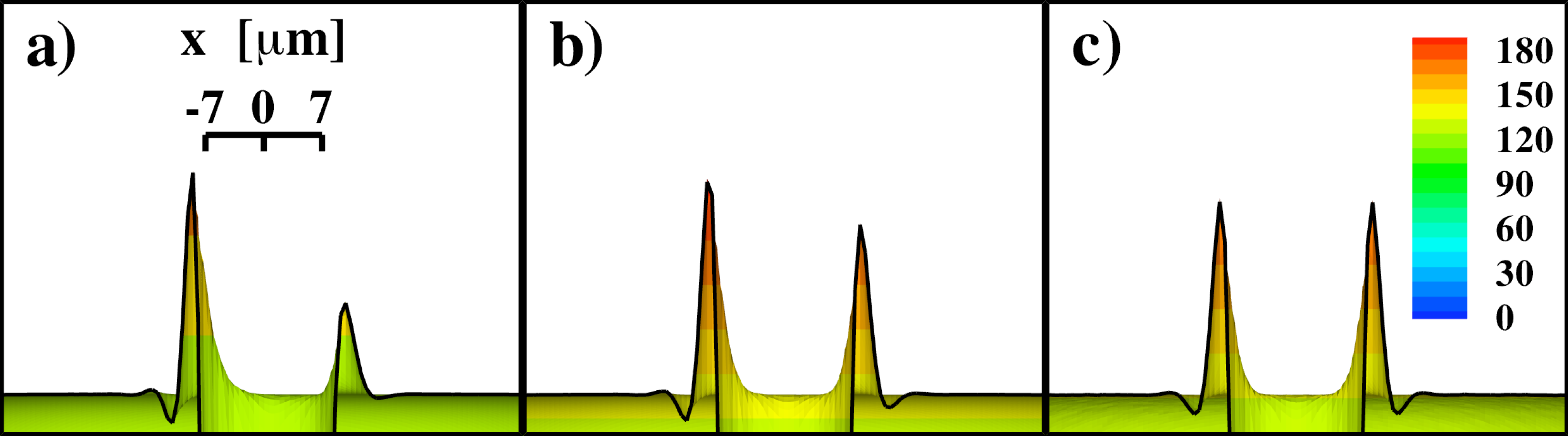}
\end{center}
\caption{(Color online) Perturbation of the steady state photon
  density $|\psi_C (\vect{r})|^2$ (side view) and the $y=0$ cut of the
  profile, $|\psi_C (x,y=0)|^2$ (solid black line), close to the
  defect, for values of the pump power asymptotically large ($c_s/v_f
  \to \infty$), where the residual drag force of Fig.~\ref{fig:dragf}
  does not have any appreciable variation.  The pump is shined on the
  cavity at a momentum $k_p=1~\mu$m$^{-1}$, and the three panels, (a),
  (b), and (c), correspond to increasing values of the polariton
  lifetime: $\hbar/\kappa_X=\hbar/\kappa_C=1$~ps (a), 3~ps (b) and
  120~ps (c). In each panel the intensity of the pumping laser has
  been adjusted in order to give the same polariton density. Note that
  in this limit, the defect induces an asymmetric perturbation of
  $|\psi_C (\vect{r})|^2$ only ahead and behind the defect, similarly
  to the left bottom panel of Fig.~\ref{fig:profi}. The asymmetry
  disappears for perfect cavities.}
\label{fig:peaks}
\end{figure}
\begin{figure}
\begin{center}
\includegraphics[width=1.0\linewidth,angle=0]{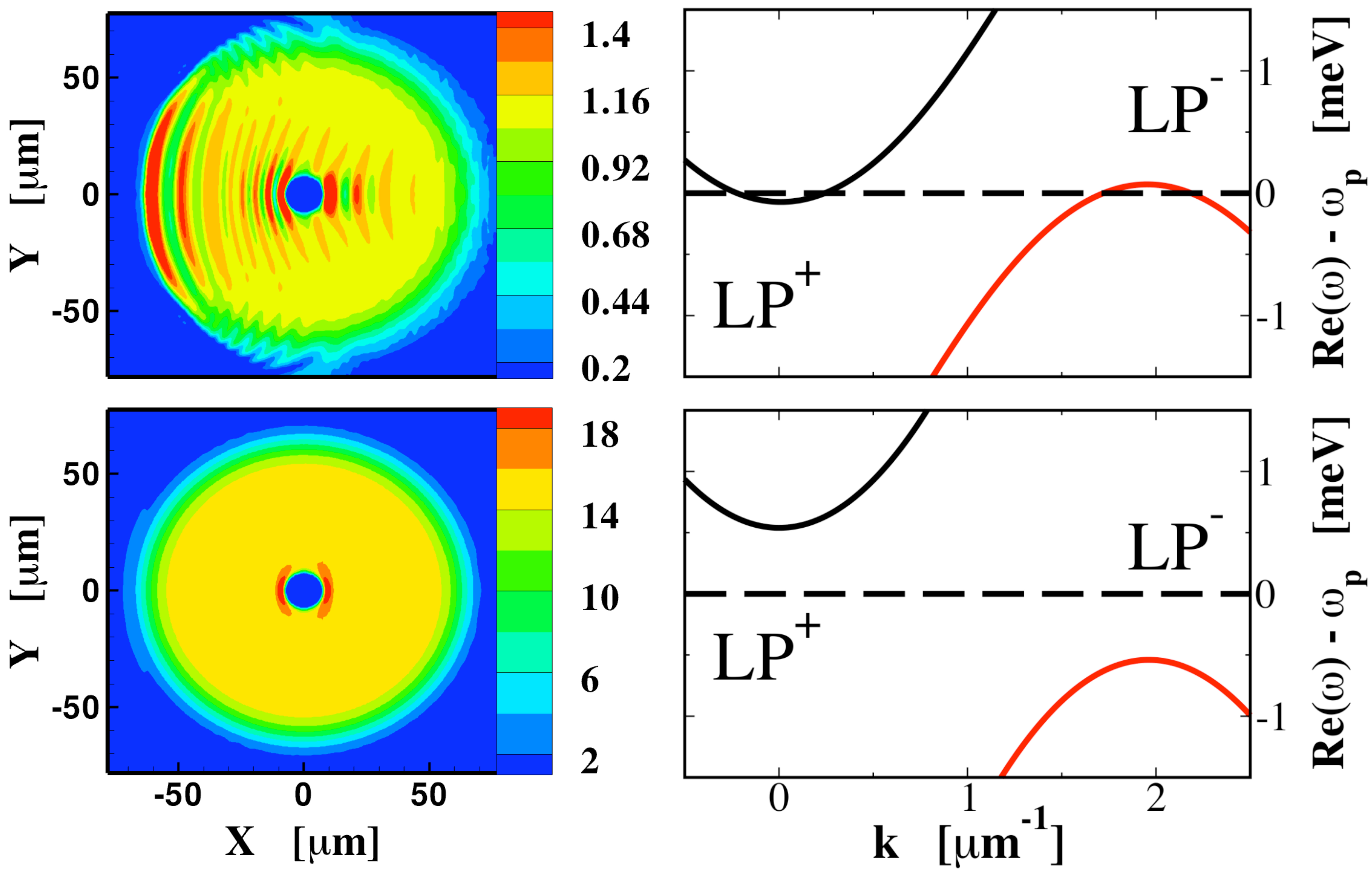}
\end{center}
\caption{(Color online) Steady state photon density profile, $|\psi_C
  (\vect{r})|^2$ (left panels) and corresponding quasi-particle
  excitation spectrum~\eqref{eq:spect} (right panels) evaluated in the
  homogeneous case for the same system parameters of the left
  panels. Here, $k_p=0.9~\mu$m$^{-1}$, $\hbar/\kappa_C=7$~ps,
  $\hbar/\kappa_X=120$~ps, and the values of the pump power correspond
  to a state where scattering is allowed in the linear approximation
  at $c_s/v_f=0.54$ (top panels) and no quasi-particle excitation in
  allowed in the linear approximation at $c_s/v_f=1.49$ (bottom
  panels). Both pump strength values considered here corresponds to
  photon densities on the upper branch of the bistability curve.}
\label{fig:profi}
\end{figure}

\begin{figure}
\begin{center}
\includegraphics[width=1.0\linewidth,angle=0]{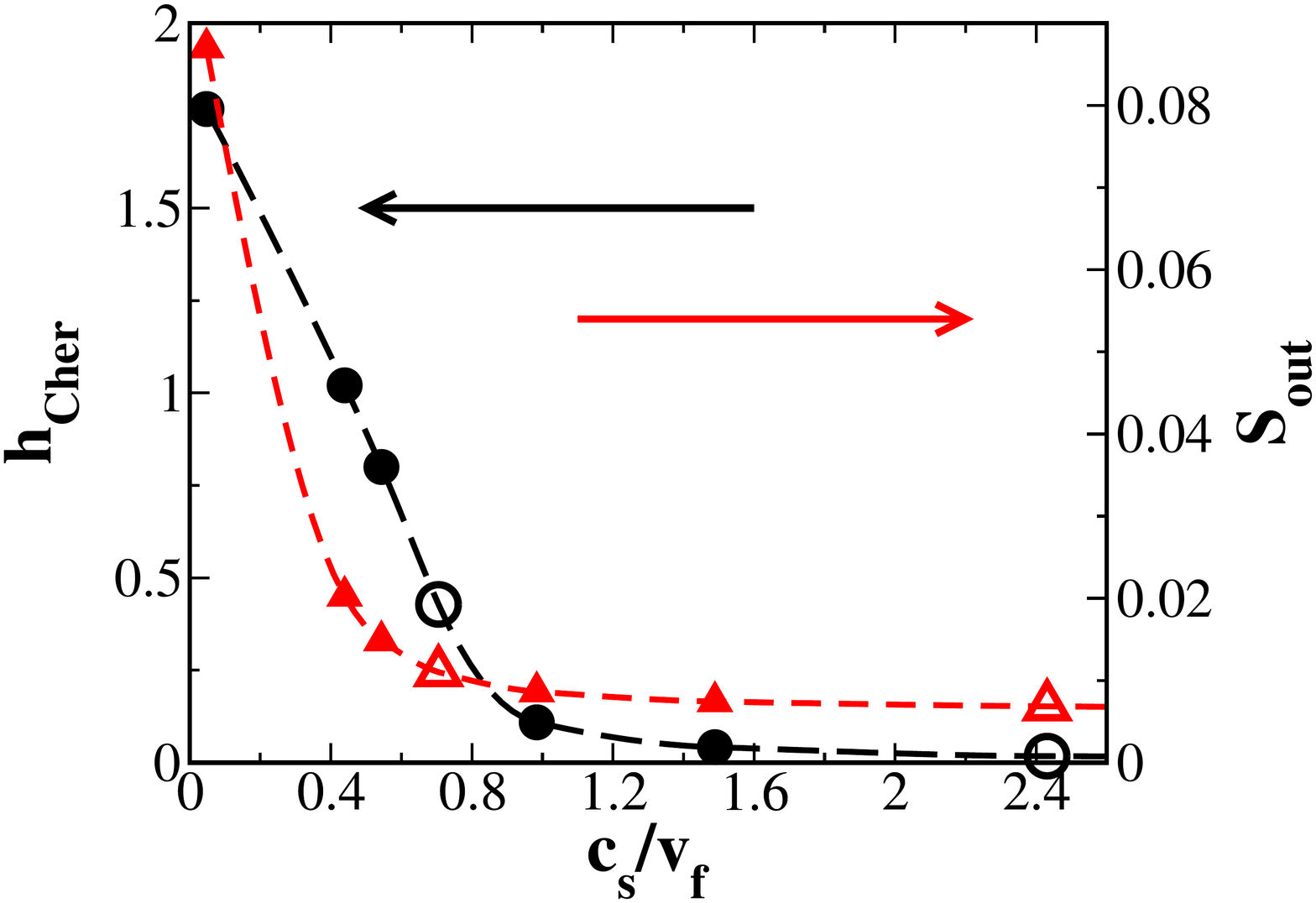}
\end{center}
\caption{(Color online) Height of the Cherenkov waves
  $h_{\text{{\u C}er}}$ (black circles) and percentage of scattered
  particles $S_{out}$ (red triangles) as defined in
  Sec.~\ref{sec:model}, plotted as a function of the normalized
  density $c_s/v_f$. The parameters are the same as the ones of
  Fig.~\ref{fig:profi}: $k_p=0.9~\mu$m$^{-1}$, $\hbar/\kappa_C=7$~ps,
  $\hbar/\kappa_X=120$~ps. The empty symbols corresponds to the values
  $c_s/v_f=0.54$ and $c_s/v_f=1.59$ chosen to plot respectively the
  top and bottom panels of Fig.~\ref{fig:profi}.}
\label{fig:heigh}
\end{figure}

\begin{figure}
\begin{center}
\includegraphics[width=1.0\linewidth,angle=0]{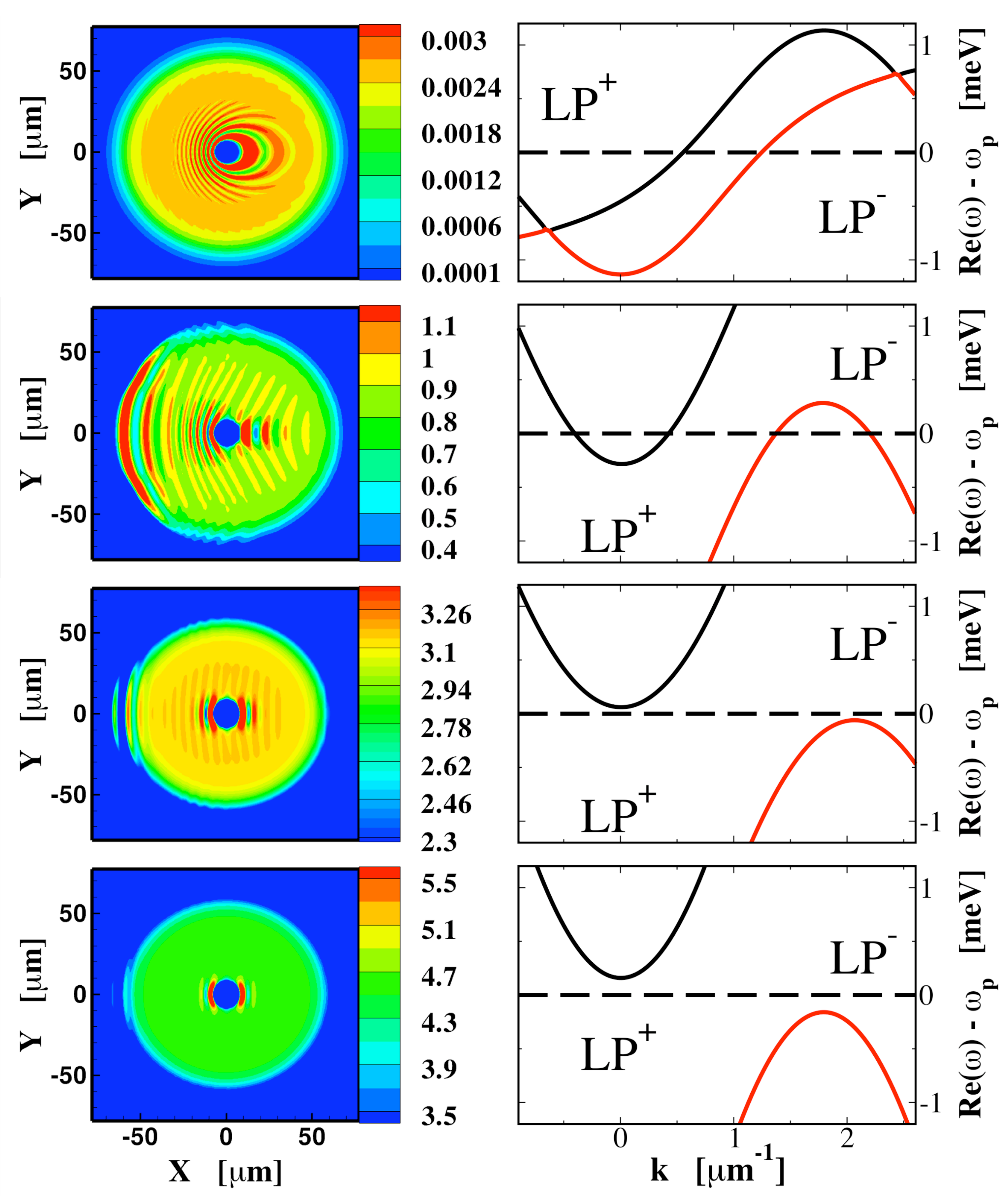}
\end{center}
\caption{(Color online) Steady state photon density profile, $|\psi_C
  (\vect{r})|^2$ (left panels) and corresponding quasi-particle
  excitation spectrum (right panels) for $k_p=1~\mu$m$^{-1}$,
  $\hbar/\kappa_C=\hbar/\kappa_X=7$~ps, and values of the pump power
  giving $c_s/v_f=0.05$ (first row), $c_s/v_f=0.7$ (second row),
  $c_s/v_f=1.25$ (third row) and $c_s/v_f=1.5$ (fourth row). Note in
  that in the third panel, while in the linear approximation no
  scattering and therefore no Cherenkov waves are allowed, the
  full solution of the problem still allows scattering, as Cherenkov
  waves are observed in the profile of the corresponding
  left panel.}
\label{fig:waves}
\end{figure}

\section{Results}
\label{sec:resul}
%
We now turn to the non-perturbative numerical analysis of the
Gross-Pitaevskii equation~\eqref{eq:model} and the analysis of the
behaviour of the superfluid properties of the resonantly driven
polariton system in the pump-only regime when changing the laser pump
strength. In particular, by fixing the pump momentum $\vect{k}_p$ and
energy $\omega_p$, we evaluate the dependence of the drag
force~\eqref{eq:dragf}, the height of the Cherenkov waves
$h_{\text{{\u C}er}}$ and the percentage of scattered particles
$S_{out}$, on the the steady-state average density of polaritons,
$n_{LP} = \int d\vect{r}|\psi_{LP} (\vect{r})|^2/\Omega$, pumped into
the cavity --- here, $\Omega$ is the circle pumping area we are
averaging the density over. Rather than as a function of $n_{LP}$, we
present the results as a function of the ratio between sound and
polariton fluid velocities, $c_s/v_f$, defined in
Eqs.~\eqref{eq:fluid} and~\eqref{eq:sound}.

We plot in Fig.~\ref{fig:dragf} the drag force exerted on the
polariton fluid by the defect~\eqref{eq:defec} in the $x$-direction,
${F_d}_x$~\eqref{eq:dragf}, as a function of the ratio
$c_s/v_f$~\eqref{eq:sound}, for different values of the pump
wavevector $\vect{k}_p$. We find that the drag force decreases fast as
a function of the polariton density, when $c_s/v_f< 1$, while, for
finite values of the polariton lifetime, reaches a finite asymptotic
value, a \emph{residual drag force}, at large densities, when $c_s/v_f
\gg 1$. While for conservative superfluid systems the drag force has a
threshold-like behaviour and in particular, for perturbatively weak
delta-like defects, is finite only for superfluid velocities above the
speed of sound~\cite{astrakharchik04}, we now observe a smooth
crossover as a function of $c_s/v_f$. In addition, as shown in the
inset of Fig.~\ref{fig:dragf}, we find that the residual drag force
vanishes only in the limit of perfect microcavities,
$\kappa_X=\kappa_C \to 0$, i.e., for infinitely long-lived
polaritons. Note that this recovers the equilibrium limit, as the
condition $\kappa_X=\kappa_C \to 0$ automatically requires a pump
strength $f\to 0$, as the laser pump strength needed to reach a given
polariton-density in the cavity decreases with the increasing
polariton lifetime.

The two main results to be drawn from Fig.~\ref{fig:dragf} are the existence of a
cross-over instead of an abrupt transition and the appearance of the
residual drag force. Since these results have been obtained with a practically
infinite barrier ($V_d=110$~meV ), one can wonder to what extend they
remain valid for actual potential barriers which are much lower. We have repeated
the analysis for different values of $V_d$, including negative ones, and always
a cross-over occurs for finite decay rates. More interesting is the behavior of
the residual drag force: it always quenches linearly with the decay rates as
shown in the right panel of Fig.~\ref{newfig} for $k_p=0.7~\mu$m$^{-1}$ in two
cases with $V_d=-22$~meV and $V_d=-2.2$~meV. The dependence of the residual
drag force on $V_d$ is shown in the left panel for $k_p=0.7~\mu$m$^{-1}$ and
$\kappa_X=\kappa_C=0.22$~meV. Calculations have been done with negative
values of $V_d$ for numerical reasons. A clear conclusion is drawn from
Fig.~\ref{newfig}: the residual drag force is non zero as far as both barrier
height and decay rates are finite.

The origin of the residual drag force for asymptotically large values
of $c_s/v_f$ at finite polariton lifetimes can be understood in terms
of the asymmetry of the perturbation generated by the defect in the
(e.g., photonic) wave-function, $|\psi_C (\vect{r})|^2$ --- see
Fig.~\ref{fig:peaks}. While, as explained later, we observe the
disappearance of the Cherenkov waves for large enough polariton
densities, even at asymptotically large polariton densities we always
observe two small perturbations in the fluid wave-function around the
defect, one just in front and one just behind the defect in the
$x$-direction, i.e., in the polariton flow direction. We plot in
Fig.~\ref{fig:peaks} cut at $y=0$ of the wave-function around the
defect. It is clear from the definition~\eqref{eq:asymm} that, in this
limit, the drag force measures the degree of asymmetry of such two
perturbations. As the perturbation becomes symmetric around the defect
in the limit of large polariton lifetimes, the drag force vanishes. We
find that the shorter the polariton lifetime the smoother is the
crossover observed in the drag force from supersonic to superfluid
behaviour. Instead, for high quality cavities, i.e. long polariton
lifetimes, the transition from supersonic to subsonic behaviour is
sharper and the residual value of the drag force is smaller.

In Fig.~\ref{fig:dragf}, for every value of the pump momentum
$\vect{k}_p$, we also identify the region of values of $c_s/v_f$ above
which in the linear approximation the Landau condition of
Eq.~\eqref{eq:landa} cannot be satisfied, and therefore where in the
linear approximation one would not expect any drag (such values are
plotted as empty symbols). Two examples of the linear excitation
spectrum are plotted in Fig.~\ref{fig:profi} together with the
corresponding profiles evaluated beyond the linear approximation ---
Note that in order to do that the pump strength $f$ in the mean-field
equations of Sec.~\ref{sec:linea} needs to be fixed so that to give
the same average density of polaritons evaluated by solving
numerically Eq.~\eqref{eq:model}, $|\psi_{X,C}^{(0)}|^2 = \int
d\vect{r}|\psi_{X,C} (\vect{r})|^2/\Omega$. We find that, while
eventually the Cherenkov waves disappear at large enough
densities, the persistence of asymmetric perturbations around the
defect, which we ascribe to finite lifetime effects, contributes to
give a finite drag force. This is also apparent in
Fig.~\ref{fig:heigh}, where we plot the height of the Cherenkov
waves as a function of $c_s/v_f$ for the same system parameters as
Fig.~\ref{fig:profi}. Here, it is clear that the the Cherenkov
waves height is strongly suppressed for $c_s/v_f \simeq 1$, and go to
zero for $c_s/v_f \gg 1$.

In addition, in Fig.~\ref{fig:heigh} we plot the percentage of the
particles scattered by the defect. Here, like for the drag force, we
find a residual value of the percentage of scattered particles at
asymptotically high polariton densities. The difference with the drag
force is here that, even for perfect cavities when the drag goes to
zero, the percentage of scattered particles keeps retaining a residual
value (see panel (c) of Fig~\ref{fig:peaks}).

In Fig.~\ref{fig:waves} we plot the photon density profiles $|\psi_C
(\vect{r})|^2$ (left panels) for increasing pump power, obtained by
solving the time dependent Gross-Pitaevskii equation~\eqref{eq:model},
together with the spectrum of linear excitations obtained by solving
the eigenvalue problem~\eqref{eq:spect} (right panels). Left panels
show Cherenkov waves evolving smoothly from a `closed' to an
`open' shape till they disappear when the subsonic superfluid regime
is reached. In particular, the angle formed between the waves and the
propagation direction increases by increasing the polariton
density. Note also that, while qualitatively the phenomenology of the
Cherenkov waves seems to be well described by the linear
approximation theory, the transition from the supersonic to the
subsonic superfluid regime is not. In particular, we find a value
range of the pump power (like the one shown in the third row of
Fig.~\ref{fig:waves}), where no scattering and therefore no Cherenkov
waves are allowed in the linear approximation description,
while the full numerical solution of the time dependent
Gross-Pitaevskii equation~\eqref{eq:model} still allows scattering of
polaritons and the associated Cherenkov waves.

\section{Conclusions and discussion}
\label{sec:concl}

In this paper we have proposed three different ways to analyse the
superfluid properties of coherently driven polaritons in the pump-only
configuration in presence of a defect potential. We have evaluated the
drag force when the fluid passes the defect, the height of Cherenkov waves,
and the percentage of particles scattered by the
defect. By making a comparison with the linearised Bogoliubov-theory
introduced in Refs.~\cite{carusotto04,ciuti05}, we have found that,
the disappearance of the Cherenkov waves, characterising the
transition from a supersonic to a subsonic superfluid behaviour, is
not well described by the linear theory. In addition, we have found
that non-equilibrium effects which go beyond the linear approximation
cause a finite residual drag force even at asymptotically large
polariton densities, where the Landau criterion predicts that no
elementary excitation can be emitted by the defect. Only in the limit
of infinitely long polariton lifetimes, the residual drag force at
large enough densities goes to zero, recovering the equilibrium
limit. The drag force exerted on the polariton fluid by a defect, as
well as the height of Cherenkov radiation, and the percentage of
particles scattered by the defect show a smooth crossover rather then
a sharp threshold-like behaviour which is typical of superfluids
obeying the Landau criterion. We have characterised this crossover as
a function of the fluid velocity, the polariton density and the
polariton lifetime.

The three observables which we here evaluate theoretically, can in
principle be measured in current state-of-the art experiments on
semiconductor microcavities. For example, the defect can be carefully
engineered by either patterning a metal grating on the microcavity top
mirror or growing mesas in one of the mirrors~\cite{eldaif06}. This
allows having a predetermined shape and size of the defect, suitable
for a direct comparison with our theoretical analysis.  Alternatively,
the defect can be switched on and off externally by an additional
laser~\cite{amo10}. The second scheme would allow a direct comparison
between the fluid motion in presence and absence of the defect. Both
schemes are within the current experimental reach, and so with this
work we intend to motivate further experimental investigations, which
would lead to a better understanding of the novel non-equilibrium
superfluid phenomena in microcavities.

\acknowledgments We are grateful to I. Carusotto, J. Keeling,
D. Sanvitto, and L. Vi\~na for continuous stimulating
discussions. This research has been supported by the Spanish MEC
(MAT2008-01555, QOIT-CSD2006-00019) and CAM
(S-2009/ESP-1503). F.M.M. acknowledges financial support from the
programs Ram\'on y Cajal and INTELBIOMAT (ESF).


\newcommand\textdot{\.}

\end{document}